\documentclass[aps,prl,twocolumn,groupedaddress,superscriptaddress]{revtex4-1}
\usepackage{graphicx}
\usepackage{amsmath}
\usepackage{epstopdf}
\usepackage{float}
\usepackage{xcolor}
\usepackage{appendix}
\begin{document}
%\preprint{}
\title{Cooling a Fermi gas with three-body recombination near a narrow Feshbach resonance}

\author{Shuai Peng}
%\thanks{These authors contributed equally to this work.}
\affiliation{School of Physics and Astronomy, Sun Yat-sen
    University, Zhuhai, Guangdong, 519082, China}

\author{Haotian Liu}
%\thanks{These authors contributed equally to this work.}
\affiliation{School of Physics and Astronomy, Sun Yat-sen University, Zhuhai, Guangdong, 519082, China}

\author{Jiaming Li}
%\thanks{These authors contributed equally to this work.}
\email[]{lijiam29@mail.sysu.edu.cn}
\affiliation{School of Physics
    and Astronomy, Sun Yat-sen University, Zhuhai, Guangdong, 519082,
    China}
\affiliation{Center of Quantum Information Technology,
    Shenzhen Research Institute of Sun Yat-sen University, Shenzhen,
    Guangdong, China 518087}
\affiliation{Guangdong Provincial Key
    Laboratory of Quantum Metrology and Sensing, Zhuhai, Guangdong,
    519082, China}

\author{Le Luo}
\email[]{luole5@mail.sysu.edu.cn}
\affiliation{School of Physics and
    Astronomy, Sun Yat-sen University, Zhuhai, Guangdong, 519082, China}
\affiliation{Center of Quantum Information Technology, Shenzhen
    Research Institute of Sun Yat-sen University, Shenzhen, Guangdong,
    China 518087}
\affiliation{Guangdong Provincial Key Laboratory of
    Quantum Metrology and Sensing, Zhuhai, Guangdong, 519082, China}
%\homepage[]{Your web page}
%\thanks{}
%\altaffiliation{}

%--------------------------------------------------------------------------------------%

\date{\today}
\begin{abstract}
    Three-body recombination is a phenomenon common in atomic and molecular collisions,
    producing heating in the system. However, we find the cooling effect of the
    three-body recombination of a $^6$Li Fermi gas near its $s$-wave
    narrow Feshbach resonance. Such counter-intuitive behavior is explained as follows, the threshold energy of the quasi-bounded
    Feshbach molecule acts as the knife of cooling, expelling the
    scattering atoms with selected kinetic energy from the trap. When
    the threshold energy happens to be larger than $3/2 k_B T$, each lost atom in the three-body
    recombination process has more than $3 k_B T$ energy which results in
    cooling. The best cooling is found with the threshold energy set at about 3$k_B T$, consistent with a
    theoretical model. The three-body recombination induced cooling raises potential applications for
    cooling complex atomic systems.
\end{abstract}
%\pacs{313.43}
\maketitle

%--------------------------------------------------------------------------------------%
%\section{Introduction}
\textit{Introduction}. Cooling method promotes the studies of ultracold quantum physics~\cite{Chu1985PRL55.48, Lett1988PRL61.169}, for example, the understanding and implementation of evaporative cooling boosts the production of Bose-Einstein condensation (BEC)~\cite{Anderson1995Science269.5221,Davis1995PRL75.3969} as well as degenerate Fermi gases~\cite{OHara2002Science298.5601,DeMarco1999Science285.1703,Truscott2001Science291.5513}.
Thus, novel cooling techniques are very important and encouraged for ultracold physics~\cite{Marco2019Science363.853,Mathey2009PRA80.030702,Li2016PRA93.041401,Dogra2019PRL123.020405,DeSalvo2017PRL119.233401}.
Evaporative cooling, which lowers the temperature by removing the atoms with higher energy, is commonly used to cool the atomic gases to an ultracold regime~\cite{Luiten1996PRA53.1}.
Similarly, a controllable and energy selective atom loss is one of the key points in most cooling schemes.

Feshbach resonance, which happens when the molecular bound state in a closed scattering channel approaches the scattering state in a open channel, is widely applied to control the interaction between atoms in ultracold experiments~\cite{Chin2010RMP82.1225}.
In the Bardeen-Cooper-Schrieffer (BCS) side of a narrow Feshbach resonance, the threshold energy of the molecule $E_t$ is positive, and a quasi-bound molecule temporarily formes when the relative kinetic energy of the two scattering atoms matches the threshold energy~\cite{Kohler2006RMP78.1311}.
Once this quasi-bound molecule collides with another incoming atom, a deeper-bound molecule generates through indirect three-body recombination~\cite{Li2018PRL120.193402, Waseem2019PRA99.052704}.
At the same time, the enormous stored energy between the two molecular states will be released to the atoms and result in atom loss.
%Because of the energy-releasing and atom loss, three-body recombination often associates with heating in ultracold systems.
Although three-body recombination should be suppressed in most experiments, it is surprisingly suggested to cool Bose gases most recently.
In Ref.~\cite{Schemmer2018PRL121.200401}, three-body recombination successfully lowers the temperature of a one-dimensional Bose gas.
In their system, the temperature is lowered to a quarter. But because of the large three-body atom loss, the gas is hard to reach more degenerate.
Following their work, Ref.~\cite{Dogra2019PRL123.020405} proposes that three-body recombination can make a three-dimensional (3D) homogeneous Bose gas more degenerates if the initial conditions are properly set.

In this paper, we experimentally demonstrate three-body recombination can be applied to cool a 3D harmonically trapped $^6$Li Fermi gas.
A magnetically tunable three-body recombination near the $s$-wave narrow Feshbach resonance of $^6$Li is used.
By precisely regulating the strength of the three-body process and threshold energy of the quasi-bound molecule $E_t$, we find the three-body recombination not only achieves the temperature reduction but also increases the phase-space density, which is important for ultracold quantum gases.
%The best cooling happens when $E_t\simeq3 k_B T$, and the largest phase-space density is obtained at $E_t\simeq9 k_B T/2$.
%Here, $k_B$ is the Boltzmann constant and $T$ is the initial temperature of the gas.
%These measurements are in good agreement with theoretical predictions.
The temperature dependence of this three-body cooling is also studied with varying initial temperatures. The results turn out the cooling becomes larger when the temperature is lower. And the cooling speed of three-body collisional is not sensitive to the initial temperature or trapping frequency. %The is another important feature of this novel collisional cooling, which may inspire the cooling of many other atomic species.

%%%%%%%%%%%%%%%%%%%%%%%%%%%
\begin{figure}[htbp]
    \begin{center}
        \includegraphics[width=\columnwidth, angle=0]{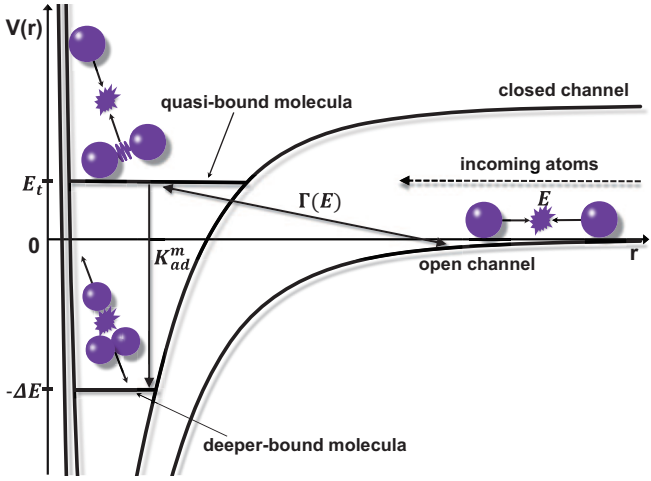}
        \caption{Schematic of indirect three-body recombination near a $s$-wave narrow Feshbach resonance. The black curves are the potentials of the open and closed scattering channels, respectively. A quasi-bound molecular state with threshold energy of $E_t$ is shown in the closed channel.
        A deeper-bound molecule has much lower energy than the quasi-bound one is also distributed in the closed channel, which generates from the upper state with a rate of $K^m_{ad}$ through the three-body process. The quasi-bound molecule decays back to the open channel with a rate of $\Gamma(E)$.
        } \label{p:Fig.1}
    \end{center}
\end{figure}
%%%%%%%%%%%%%%%%%%%%%%%%%%%
\textit{Experiments and results}. The two lowest-energy hyperfine ground-state mixtures of ultracold $^6$Li Fermi gases have a $s$-wave narrow Feshbach resonance around 543.3 G with a resonance width of 0.1 G.
We prepare the ultracold atom gases to follow the procedure described in Ref.~\cite{Chen2021PRA103.063311, Li2018PRL120.193402}.
As Fig.~\ref{p:Fig.1} illustrated, because of the narrow resonance width, the relative kinetic energy of two scattering atoms with a temperature of $T$ is comparable with the $E_t$, which means a quasi-bound molecule forms even in the BCS regime without an extra energy injection.
The quasi-bound molecule will decay back into the initial scattering channel or decay into a deeper-bound molecule state through indirect three-body recombination.
In the three-body process, if the energy loss per particle is larger than the sum of the releasing energy of the deep bound molecule and the average energy per atom of the gas $3 k_B T$, the total energy of the residual trapped atoms is reduced.  Here, $k_B$ is the Boltzmann constant.
This cooling mechanism of three-body recombination is inspired by the collisional cooling method in Ref.~\cite{Mathey2009PRA80.030702,Nuske2015PRA91.043626}.
We replace the two-body inelastic decay of the quasi-bound molecule with a three-body process. The reason is obvious that the other open scattering channel with lower energy is absent in the two lowest-energy hyperfine ground-state mixtures of $^6$Li.
It is noticed that the quasi-bound molecule will decay back to the initial open channel without atom or energy changing, which can be treated as a kind of elastic collision and helps the gas achieve thermalization~\cite{Mathey2009PRA80.030702}.
In indirect three-body recombination, the three-body process strongly relates to the two-body recombination. In another word, if the system is absent of the direct three-body recombination, the rate of three-body recombination is lower than the formation of the quasi-bound molecule. Therefore, $\Gamma_{tot}=\Gamma_0+\Gamma(E)>\Gamma(E)\gg\Gamma_0\propto K^m_{ad}$, where $\Gamma_{tot}$ describes the total linewidth.

The three-body loss is $\dot{N_3}=-L_3\int n^3(x,y,z)dxdydz$, and the corresponding potential energy loss rate is $\dot{E_{3p}}=-L_3\int n^3(x,y,z) U(x,y,z)dxdydz$. Thus, the lost potential energy per lost atom is $\dot{E_{3p}}/\dot{N_3} =1/3 U(\sigma_x,\sigma_y,\sigma_z) $. Here, $\sigma_{x,y,z}$ is the Gaussian width of $n(x,y,z)$.
According to the Virial theorem, the average kinetic energy per particle $\langle E_k\rangle=3/2 k_B T$ is equal to the average potential energy $\langle E_p\rangle= U(\sigma_x,\sigma_y,\sigma_z)$ in a harmonically trapped gas~\cite{LuoLePHDthesis, VirialTheoremExplain}.
Then, $\dot{E_{3p}}/\dot{N_3} =1/3\langle E_p\rangle$, which indicates the average of the potential energy loss in a three-body collision is $3 k_B T/2$.
Therefore, the Boltzmann equation of the collisional cooling approaches to the 3D homogeneous gas case\cite{Mathey2009PRA80.030702}. And it further simplifies into the rate equations of density and temperature (more detail presents in the supplementary materials (SM))
\begin{equation} \nonumber
\label{eq:1}
\frac{\partial n}{\partial t} = - L_3 n^3,
\end{equation}
\begin{equation}
\label{eq:2}
\frac{\partial T}{\partial t} = - L_3 n^2(\frac{E_{t}}{3 k_B}-\frac{T}{2}),
\end{equation}
with three-body loss rate~\cite{Li2018PRL120.193402, Chen2021PRA103.063311, Hazlett2012PRL108.045304}
\begin{equation}
\label{eq:3}
L_3(E_t,T)=3 K^m_{ad}(\sqrt{2}\lambda_T)^3e^{-E_t/k_B T}.
\end{equation}
Here, $ K^m_{ad}$ is the formation rate of the deeper-bound molecules, thermal wavelength $\lambda_T=\sqrt{2 \pi \hbar^2/m k_B T}$, and threshold energy $E_t=2\mu_B(B-B_0)$ also expresses the binding energy of the Feshbach molecule ($\mu_B$ is the Bohr magnetron).
%%%%%%%%%%%%%%%%%%%%%%%%%%%
\begin{figure}[htbp]
    \begin{center}
        \includegraphics[width=\columnwidth, angle=0]{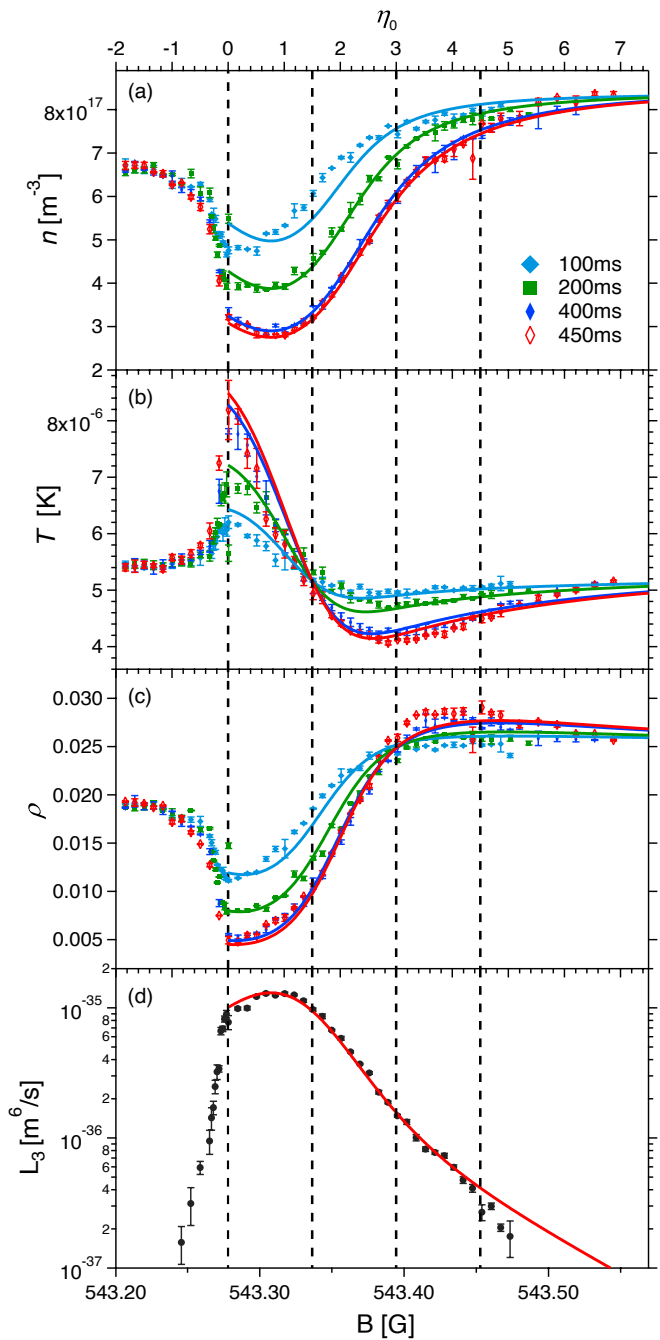}
        \caption{Experiment data and theoretical simulation of the three-body cooling with $T(0)=5.1 \mu K$. (a),(b),(c) and (d) show the changing of atom density $n(x,y,z,t)=N(t)/\sigma_x \sigma_y \sigma_z (2\sqrt{3}\pi)^{3/2}$, $T(t)$, phase-space density $\rho=(2\pi\hbar)^3 n(0,0,0)/(2\pi mk_B T)^{3/2}$, and $L_3$, respectively.
        The horizontal axis normalizes as the initial temperature $\eta_0 = E_{t}/k_B T(0)$. The black dashed lines are guiding for the $\eta_0 =0,3/2,3,9/2$, respectively. Solid curves in (a,b,c) are the theory calculation with Eq.~\ref{eq:10}. The red solid curve in (d) is the fitting results of Eq.~\ref{eq:3} with $K^m_{ad}=1.691\pm8 \times10^{-16}m^3/s$. Notably, the tendency of $L_3$ relates to the $T$ and $n$. Thus, we should the their impact to get the best L3 fitting.
        Resonance magnetic field $B_0=543.270$ G. Initial atom number $N(0)=4\times10^5$ per spin state (raw data is present in the SM), and  $T/T_F(0)=1.5$.
        }    \label{p:0.006U0}
    \end{center}
\end{figure}
%%%%%%%%%%%%%%%%%%%%%%%%%%%
Eq.~\ref{eq:1} can be solved analytically,
\begin{equation}\nonumber\label{eq:10}
    \frac{1}{n^2(t)}=2L_3(B)t+\frac{1}{n^2(0)},
\end{equation}
\begin{equation}
T(t)=2\{ \frac{E_{t}}{3k_B}-[\frac{E_{t}}{3k_B}-\frac{T(0)}{2}][\frac{n(0)}{n(t)}]^{1/2}\}.
\end{equation}
As \ref{eq:10} shows, when $E_t=3 k_B T(0)/2$, the temperature will keep constant during the cooling. In this way, the leaving atom brings average energy of $3 k_B T$ from the gas, which is the sum of average kinetic energy and potential energy.
This feature is tested in our system, as shown in  Fig.~\ref{p:0.006U0} (b).
It should be noticed that the free evaporative cooling effect in the BCS side of $^6$Li narrow Feshbach resonance is very small due to the ultra-narrow resonance width and tiny $s$-wave scattering length (for more detail seeing the SM). Another reason is that the energy difference $\Delta E$ between the two molecular states is supposed to be much larger than the trap depth, which results in the recoil kinetic energies of the three scattering atoms after collision are larger too.
They will leave the trap quickly without energy exchanging collisions.
So, the measured $T$ in Fig.~\ref{p:0.006U0} (b) is almost resulting from the three-body recombination, which is very important in our experiment.

When $\eta_0$ approaches zero, the atoms leave the trap to take less energy (between $\langle E_p\rangle$ and $3k_B T $), and the heating due to the three-body recombination becomes stronger. As $L_3$ increases, the $n$ and $\rho$ both decrease. We notice that the simulation model explains the experiment very well, which coincident interprets that the temperature increasing of the three-body recombination is not because of the huge energy-releasing ($~\Delta E$ in Fig.~\ref{p:Fig.1}) instead is coming from the energy selectively expelling of the lower energy atoms.
This may be inconsistent with previous intuitive thinking of the three-body heating but conforms to the weak evaporative cooling as discussed in the front paragraph.

Once $E_{t}$ exceeds $3 k_B T(0)/2$, the leaving atoms carry more energy ($>3 k_B T$) away from the gas, which leads to cooling.
As shown in Fig.~\ref{p:0.006U0} (b), the best cooling rate is at $\eta_0\sim 3$, and the temperature reaches $4 \mu K$ in 450 ms.
It notable that the $\rho$ keeps constant without increasing. But after this point, the $\rho$ starts to climb.
The reason is that $L_3$ decreases and $\dot{E}/\dot{N}$ increases which leave as many as lower energy atoms in the trap.
The $\rho$ reaches a maximum when $\eta_0 \sim 9/2$. This finding agrees with our numerical simulation.

%%%%%%%%%%%%%%%%%%%%%%%%%%%
\begin{figure}[htbp]
    \begin{center}
        \includegraphics[width=\columnwidth, angle=0]{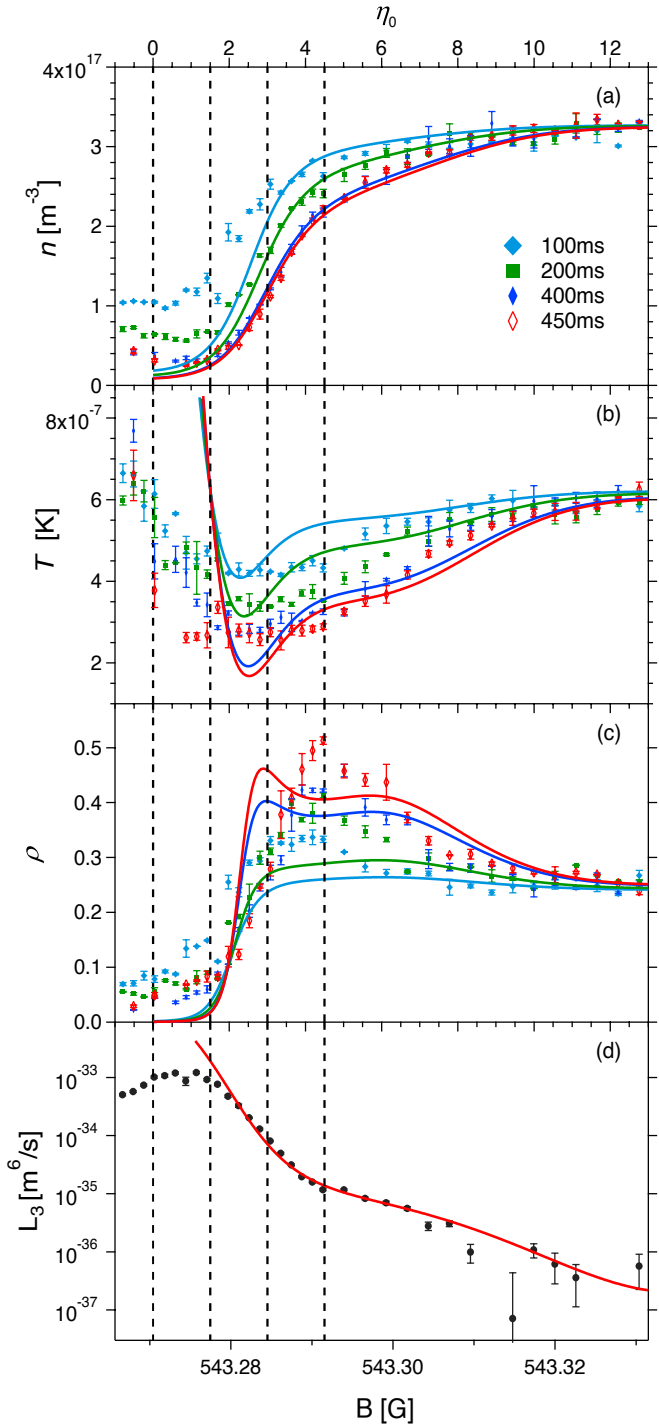}
        \caption{Experiment data and theoretical simulation of the three-body cooling with $T(0)=0.6 \mu K$. (a),(b),(c) and (d) show the changing of $n(x,y,z,t)$, $T(t)$, $\rho$, and $L_3$, respectively.
        Horizontal axis is normalized as the initial temperature. The black dashed lines are guiding for $\eta_0 =0,3/2,3,9/2$, respectively. Solid curves in (a,b,c) are the theory calculation with Eq.~\ref{eq:10}. The red solid curve in (d) is the fitting results of Eq.~\ref{eq:3} with $K^m_{ad}=2.02\pm2 \times10^{-15}m^3/s$. Notice that, as the $\rho$ increases further, the kink around $\eta_0=9/2$ of the $L_3$ becomes shaper. Initial atom number $N(0)=1.7\times10^5$ per spin state.
        }    \label{p:0.0005U0}
    \end{center}
\end{figure}
%%%%%%%%%%%%%%%%%%%%%%%%%%%

%SEq.~\ref{eq:10} shows, with a set $\eta_0$, the lower temperature limit is largely coming from the faster three-body loss rate, which estimates as $L_3\propto 1/T^{3/2+2.2}$~\cite{TempScalingLaw}.

The cooling scheme presented here shows several properties
significant different from the evaporative cooling. First, through
three-body recombination, the atoms with relative kinetic energy of
$E_t$ are selectively removed with a three-body loss rate
$L_3\propto 1/T^{3/2} \text {exp}[{-E_t/k_B T}]$. This allows a
higher cooling rate when the temperature is lower, which is opposite
to the evaporative cooling where the cooling efficiency is usually
reduced when the temperature is reduced. As shown in
Fig.~\ref{p:0.0005U0}, when $T(t=0)=0.6 \mu K$, where the initial
$T/T_{F}(t=0)$ of the gas is about 0.97, where $T_F=(6 N)^{1/3}\hbar
\bar{\omega}$ is the Fermi temperature. The $\rho$
increases from 0.25 to 0.5 at $\eta_0 = 4.5$, and the lowest $T/T_F$
is 0.79, which exhibits better cooling efficiency compared with the
higher temperature case in Fig.~\ref{p:0.006U0}. Such a temperature
dependence could be used to explain the observations in
Ref.~\cite{Hazlett2012PRL108.045304}, where a weak cooling around
543.45 G when $T(0)=3.2 \mu$K is observed, but the effect disappears
in the higher temperature.

Second, with a much lower initial temperature, the atom loss becomes much
significant, as $L_3\propto e^{-\eta}$, when $\eta$ approaches the resonance, and the atom loss
reaches an extremely high value, making any prediction without
considering how the atoms leave the trap not reliable, so the
cooling process ends in the near resonance regime.

%--------------------------------------------------------------------------------------%
%\section{Summary and discussion}
%--------------------------------------------------------------------------------------%
\textit{Discussion and conclusion}. In conclusion, a $^6$Li Fermi gas is found can be cooled to a lower
temperature with a higher phase-space density by three-body
recombination. Different from the evaporative cooling where the high
energy atoms are removed from the trap, three-body cooling
selectively expels the atoms with a certain kinetic energy related
to the scattering property. This makes the cooling efficiency not
decreased with the decrease of the cloud temperature. Accordingly,
the cost for such cooling mechanism is the stability of the magnetic
field used for the collisions resonance. In our experiment, the
effective cooling regime is only about $k_B T/2\mu_B \sim$50 mG,
which requires the stability of the magnetic field to reach at least
10 mG, around a part per million. Suggested by
Ref.~\cite{Mathey2009PRA80.030702}, collisional dependent cooling
schemes have the potential of directly cooling a thermal gas to the
deep degenerate regime. However, from an experimental point of view,
further cooling beyond our current study requires the magnetic field
has higher resolution and fast switching speed, which is absent in
in our current system~\cite{Chen2021PRA103.063311}. We expect, in
the future, three-body recombination induced cooling could be
explored for its potential application for cooling complex atomic
and molecular systems, such as dipole
molecules~\cite{Marco2019Science363.853} and mixed atomic species with rich
narrow Feshbach resonances~\cite{Wang2011PRA83.042710,Pilch2009PRA79.042718}.

%\section*{Acknowledgements}
This work is supported by the Key-Area Research and Development
Program of Guangdong Province under Grant No.2019B030330001, NSFC
under Grant No.11774436 and No.11804406, Science and Technology
Program of Guangzhou 2019-030105-3001-0035. JLi received supports
from Fundamental Research Funds for Sun Yat-sen University 18lgpy78.
LL received supports from Guangdong Province Youth Talent Program
under Grant No.2017GC010656 and Sun Yat-sen University Core
Technology Development Fund.
%-------------------------------------------------------------------------------------%

%--------------------------------------------------------------------------------------%
\newpage

\appendix
\section{Supplementary Materials}

\setcounter{equation}{0}
\setcounter{subsection}{0}
\setcounter{figure}{0}
\renewcommand{\theequation}{S.\arabic{equation}}
\renewcommand{\thesubsection}{S.\arabic{subsection}}
\renewcommand{\thefigure}{S.\arabic{figure}}

\subsection{estimation of the free evaporative cooling rate}
\begin{figure}[htbp]
    \begin{center}
        \includegraphics[width=\columnwidth]{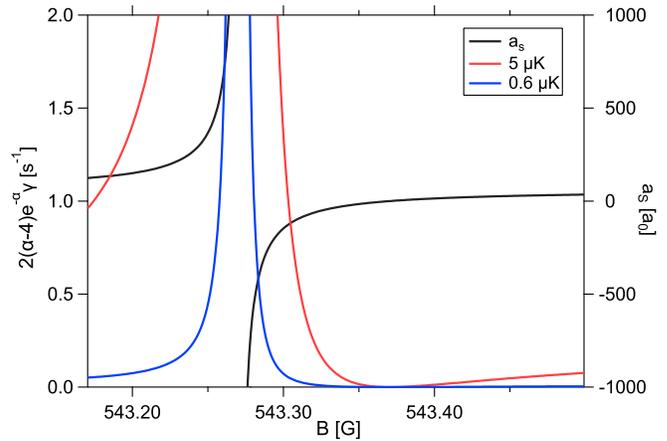}
        \caption{$s$-wave scattering length, evaporative cooling rates versus magnetic field near the narrow Feshbach resonance. $\alpha=5$. The black curve is plotting of $a_s$. The blue curve presents the $T= 5\mu K$ and the red curve is for the $T=0.6\mu K$ case.
        }    \label{p:EC}
    \end{center}
\end{figure}

The consideration of evaporative cooling is very important in our experiment. If the evaporative cooling rate is too strong, the real effect from the three-body cooling will be enlarged and incurs an overestimation.
The scattering length $a_s$ near the narrow Feshbach resonance of $^6$Li is expressed as $a_s(B)=a_{bg}(1-\Delta/(B-B_0))$, where $a_{bg}\simeq 62 a_0$ is the background scattering length, $\Delta=0.1 G$ is the resonance width and $B_0=543.270 G$ is the resonance magnetic field.
As shown in SFig.~\ref{p:EC}, $a_s$ equal to zero when magnetic field detuning $B-B_0=\Delta$, and keeps in a small range around this point, which means we can use the energy-independent $s$-wave scattering to analyze this free evaporative cooling~\cite{Luo2006NJP8.213,OHara2001PRA64.051403}.
The atom loss rate is
\begin{equation}
\dot{N}=-2(\alpha-4)\text{exp}(-\alpha)\gamma N
\end{equation}
where $\alpha= U_{trap}/k_B T$ ($U_{trap}$ is the trap depth of the optical dipole trap) and $\gamma=N \sigma m \bar{\omega}/2\pi^2k_B T$ is the collision rate with a cross section $\sigma= 4\pi a_s^2$. Here, $\bar{\omega}$ is the average geometry trapping frequency of $U_{trap}$.
%%%%%%%%%%%%%%%%%%%%%%%%%%%
%%%%%%%%%%%%%%%%%%%%%%%%%%%
We plot the evaporative atom loss rates in SFig.~\ref{p:EC}.
The atom loss rate due to the free evaporative cooling is very small and can be ignored around the zero scattering length point.
This is much different from other systems~\cite{Rem2013PRL110.163202}. The small $a_{bg}$ length and ultra-narrow width play an important role in our system. They make the free evaporative cooling rate has little contribution to the atom loss and temperature. And, in this situation, the cooling of the three-body recombination can be easily measured, or the measurements of the temperature and atom loss is the combination of three-body recombination and free evaporative cooling.
\subsection{Derivation of $\partial_t n$ and $\partial_t T$ from Boltzmann equation }
According to the previous discussion, indirect three-body recombination strongly dependent on the combination of the quasi-bound molecule. So, we derive the indirect three-body recombination from a two-body inelastic collision by replacing the inelastic decay rate $\Gamma_0$.
The differential equation of atom density and temperature is simplified from a Boltzmann equation~\cite{Mathey2009PRA80.030702}
%\begin{equation}\label{SBE1}
%\partial_t f(\vec{p},t)=-\int \frac{f(\vec{k},t)f(\vec{p},t)d^3k}{(2\pi\hbar)^2}\frac{\hbar}{m k_r}\frac{\Gamma(E)(\Gamma_0)}{(E-E_t)^2+\Gamma^2_{tot}/4}
%\end{equation}
\begin{equation}\label{SBE1}
    \partial_t f(\vec{p},t)=-\int \frac{f(\vec{k},t)f(\vec{p},t) L_2 d^3k}{(2\pi\hbar)^3},
\end{equation}
with a two-body process of
\begin{equation}
L_2=\frac{h^2}{2\pi m k_r}\frac{\Gamma(E)\Gamma_0}{(E-E_t)^2+\Gamma^2_{tot}/4}.
\end{equation}
Here, $f(\vec{k}, t)$ and $f(\vec{p},t)$ are the momentum distribution of the scattering atoms, $k_r=(\vec{k}-\vec{p})/2$ is the relative momentum.
Resonance width $\Gamma(E)$ is much larger than the thee-body decay rate of the quasi-molecular sate $\Gamma_0=3\hbar K^m_{ad} n$, which simplifies the derivation of Eq.~\ref{SBE1}.
It is noticed that the coefficient 3 inside $\Gamma_0$ describes a three-body process.
%When we derive the differential equation of atom density and temperature, $\Gamma(E)$ is assumed to be much larger than  $\Gamma_0$.
In indirect three-body recombination, for simplicity, the three-body process experiences two-body recombination and a quasi-bound molecule first. Thus, if the system is absent of the direct three-body recombination, the rate of three-body recombination $K^m_{ad}$ should be smaller than the formation of the quasi-bound molecule $K_{ad}\propto\Gamma^2(E)/(E-E_t)^2+\Gamma^2_{tot}/4$.
In our experiments, as shown in Fig.[2,3] (in the main text), the measured $L_3$ can be well fit by indirect three-body model, which satisfies that the dominated three-body loss in the BCS regime of ultracold $^6$ Li narrow Feshbach resonance is coming from the indirect process.

Because of $\Gamma_0\ll\Gamma_{tot}=\Gamma(E)+\Gamma_0\ll k_BT$, the line-shape of $L_2$ can be approached as a delta function. So, we use $E=E_{t}$ to simplify the equation. This approximation explains why three-body cooling is expelling the selected energy atoms instead of cutting off the higher energy atoms from the trap.
The partial differential of $n$ is expressed as
\begin{equation}\label{eq:15}
\begin{split}
\partial_t n=&-L_3(E_t, B)n,\\
L_3(E_t,T)=&3 K^m_{ad}(T)(\sqrt{2}\lambda_T)^3e^{-E_t/k_B T}.\\
\end{split}
\end{equation}
%$\partial_t n=-\gamma_{in}(t)n\rightarrow \partial_t T=-\gamma_{in}(t)(\frac{E_{res}}{3k_B}-\frac{T}{2})$\\
Before three-body cooling, the total energy of the atom gas is
\begin{equation}
E=3Nk_BT.
\end{equation}
After cooling, about $\Delta N$ leaves the trap, the energy of the left atoms is
\begin{equation}\label{eq:s1}
\begin{split}
E'=&3(N-\Delta N)k_BT'\\
=&E-\Delta N(E_{t}+\left< U_p \right> ).
\end{split}
\end{equation}
Then,
\begin{equation}\label{eq:s2}
\Delta T=\frac{\Delta N[3k_BT-(E_{t}+\left< U_P \right> )]}{3(N-\Delta N)k_B}.
\end{equation}
Using Eq.\eqref{eq:15}, we can derive the $\partial_t T$:
\begin{equation}\label{eq:4}
\begin{split}
\partial_t T=&\lim_{\Delta t\rightarrow 0}\frac{\Delta N[3k_BT-(E_{res}+\left< U_p \right> )]}{3Nk_B\Delta t}\\
=&-L_3\frac{[(E_{res}+\left< U_p \right>)-3k_BT]}{3k_B}.
\end{split}
\end{equation}
Plug the differential equation of total atom number
\begin{equation}\label{eq:s3}
\frac{\partial_t n}{n}=\frac{\partial_t N}{N}=\frac{1}{N}\lim_{\Delta t\rightarrow 0}\frac{-\Delta N}{\Delta t}=-L_3 n^2
\end{equation}
into Eq.~\ref{eq:4}, and get
\begin{equation}\label{eq:5}
\partial_t T=-L_3 n^2 (\frac{E_{t}}{3k_B}-\frac{T}{2}).
\end{equation}

\subsection{Raw data of atom number}
Time evolution of atom number in Fig.~2,3.
%%%%%%%%%%%%%%%%%%%%%%%%%%%
\begin{figure}[htbp]
    \begin{center}
        \includegraphics[width=\columnwidth, angle=0]{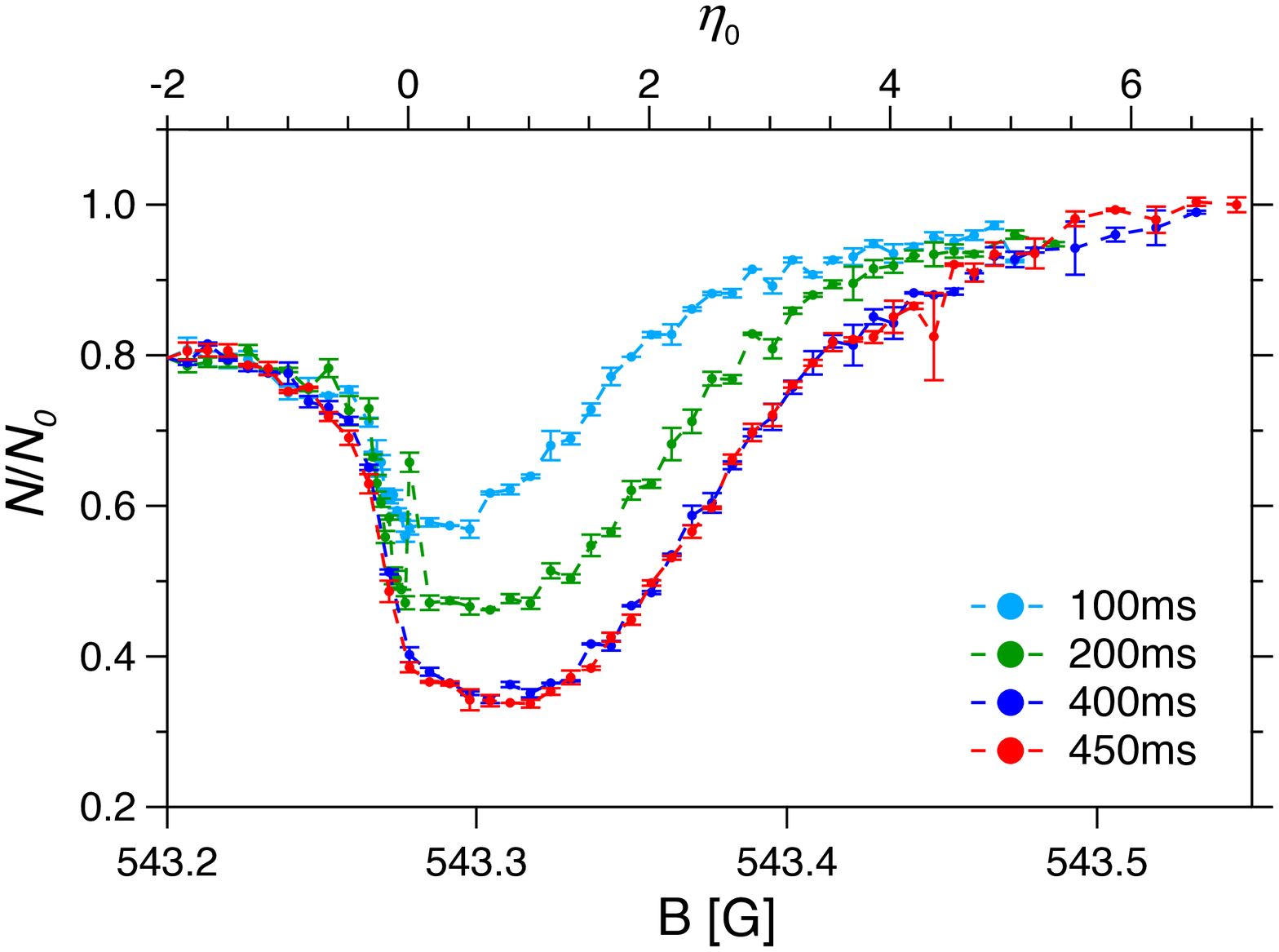}
        \includegraphics[width=\columnwidth, angle=0]{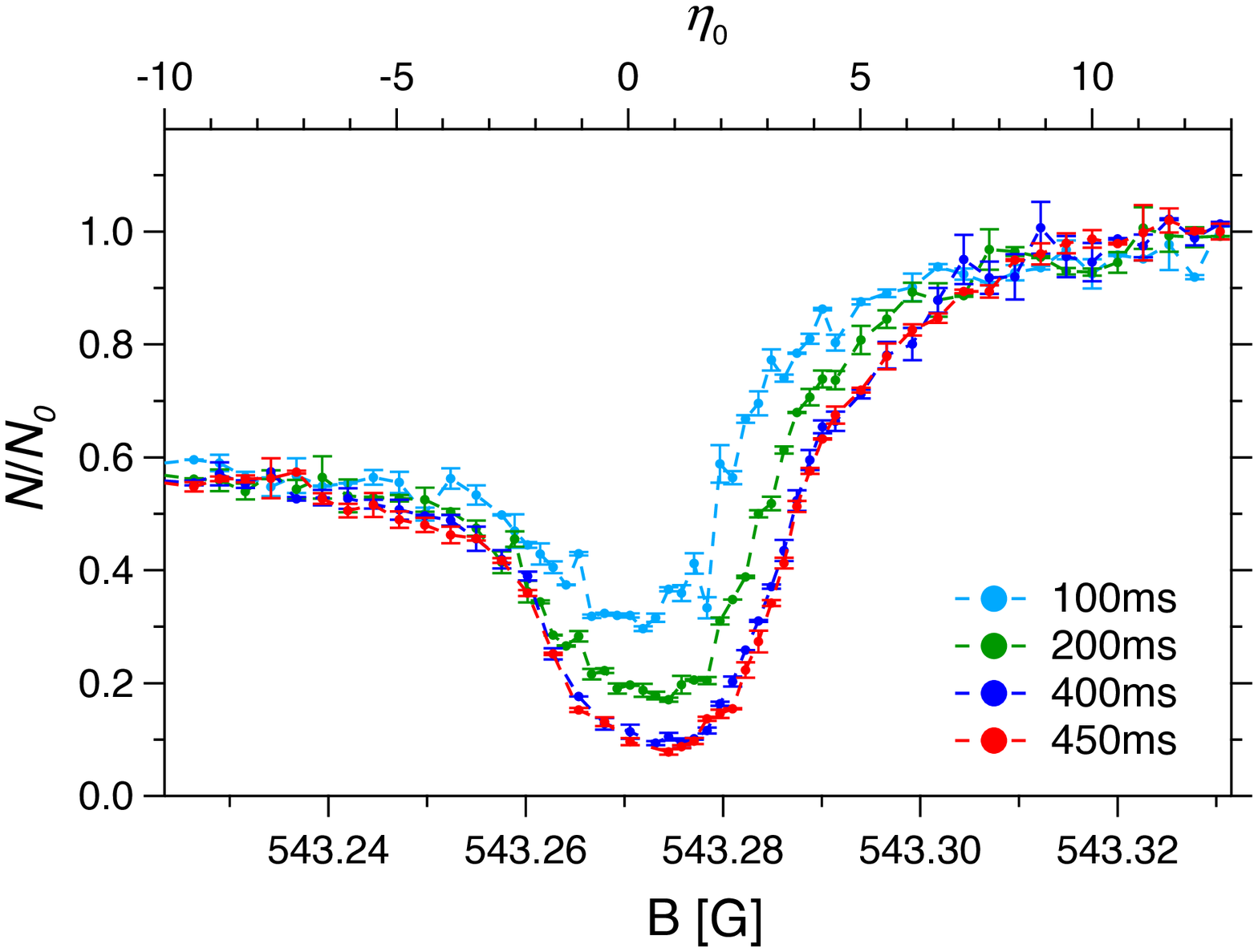}
        \caption{Upper:Atom number data of 5$\mu$K experiments. Lower: Atom number data of 0.6$\mu$K experiments.   }    \label{p:0.006U0_Number}
    \end{center}
\end{figure}
%%%%%%%%%%%%%%%%%%%%%%%%%%%

\subsection{Time evolution of the $T/T(0)$ and $\rho/\rho(0)$}
\begin{figure}[htbp]
    \begin{center}
        \includegraphics[width=\columnwidth, angle=0]{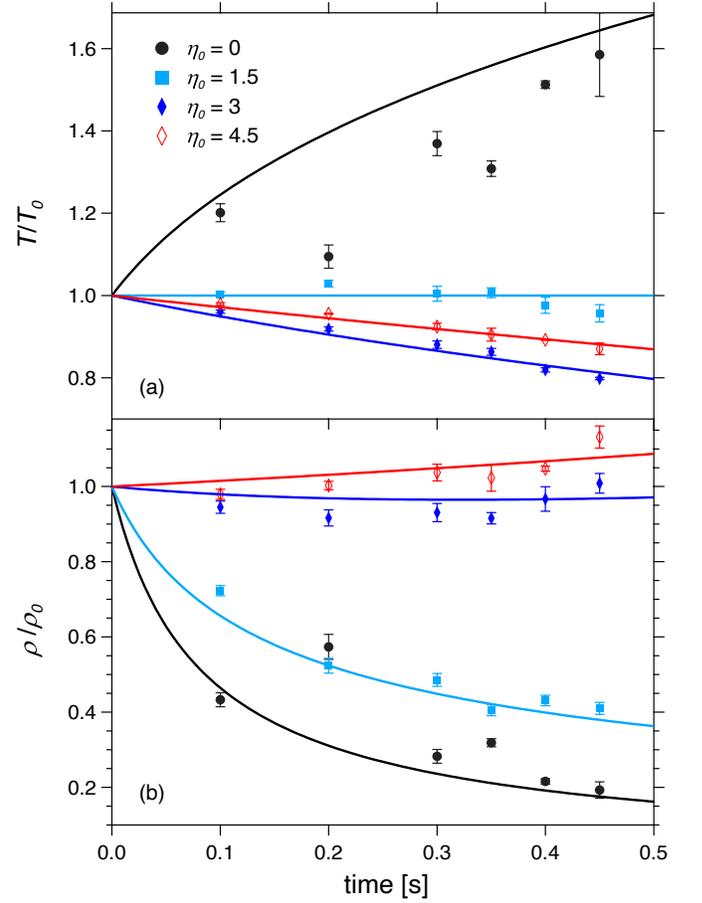}
        \caption{Time evolution of the $T/T(0)$ and $\rho$ with $\eta_0 =0,3/2,3,9/2$, when $T(0)$=5$\mu$K, respectively.
        }    \label{p:0.006U0x}
    \end{center}
\end{figure}

\begin{figure}[htbp]
    \begin{center}
        \includegraphics[width=\columnwidth, angle=0]{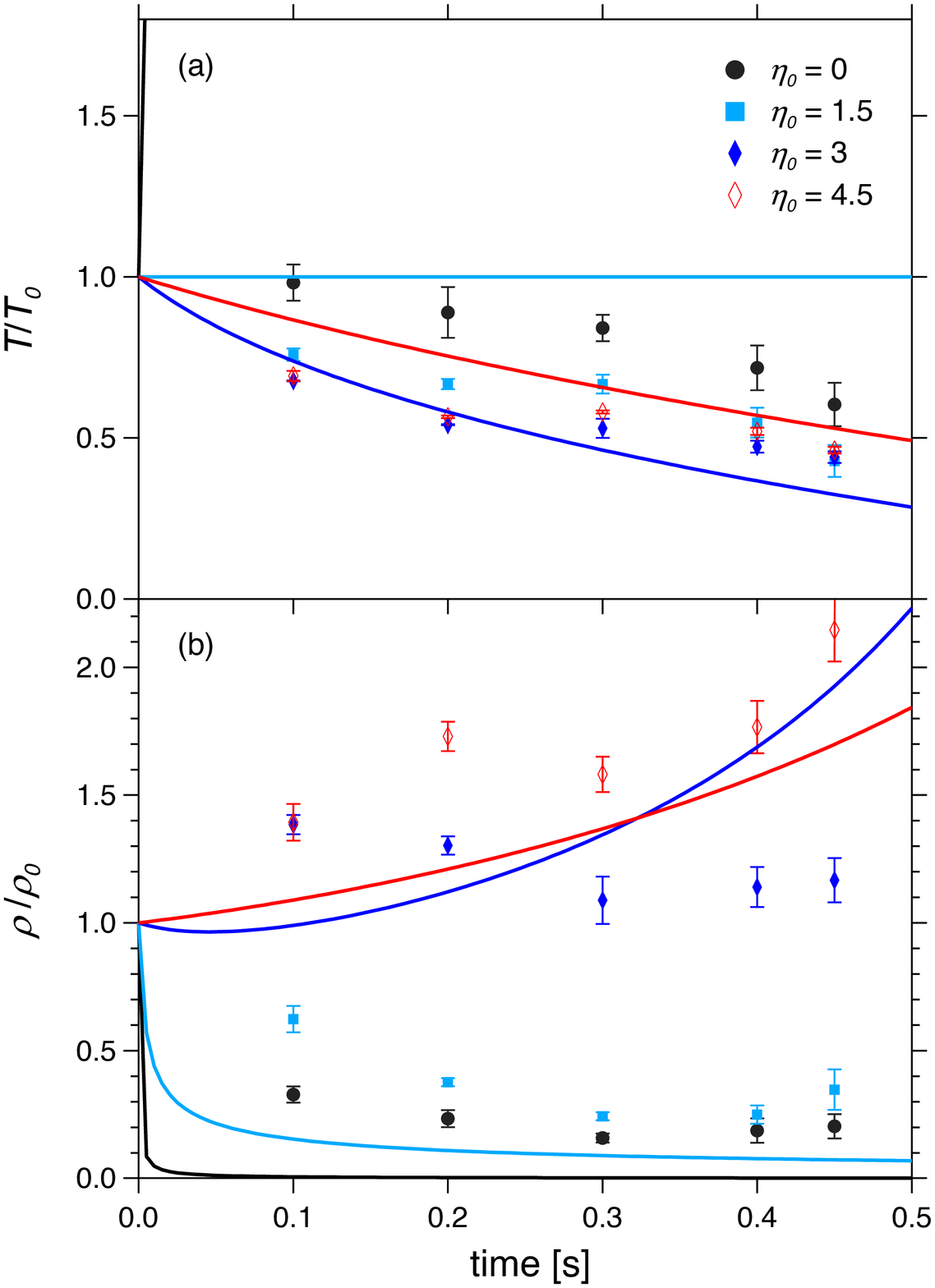}
        \caption{Time evolution of the $T/T(0)$ and $\rho$ with $\eta_0 =0,3/2,3,9/2$, when $T(0)$=0.6$\mu$K, respectively.
        }    \label{p:0.0005U0x}
    \end{center}
\end{figure}

SFig.~\ref{p:0.006U0x} and ~\ref{p:0.0005U0x} present the time evolution of temperature and phase-space density with $\eta_0$=0, 1.5, 3 and 4.5, respectively.
%%%%%%%%%%%%%%%%%%%%%%%%%%%
%%%%%%%%%%%%%%%%%%%%%%%%%%%
%%%%%%%%%%%%%%%%%%%%%%%%%%%
%%%%%%%%%%%%%%%%%%%%%%%%%%%

\subsection{Cooling effect under different temperatures}
First, we adopt the fraction of $T(t_0)$ and $T(0)$ at $t_0$ to quantify the cooling efficient.
\begin{equation}
F_T=\frac{T(t_0)}{T(0)}=2\{ \frac{\eta_0}{3}-[\frac{\eta_0}{3}-\frac{1}{2}][\frac{n(0)}{n(t_0)}]^{1/2}\}.
\end{equation}
$L_3$ increases as temperature, which results in a larger $n(0)/n(t_0)$ at a fixed time. Thus, $T(t_0)/T(0)$ becomes smaller.

In Fermi gases, we can also use the $\rho(t_0)/\rho(0)$ to characterize the cooling performance, which stands for the degenerate of the gases.
\begin{equation}
\begin{split}
F_\rho&=\frac{\rho(t_0)}{\rho(0)}=\frac{n(t_0)}{n(0)}\left[\frac{T(0)}{T(t_0)}\right]^{3/2}\\
&=\frac{n(t_0)}{n(0)}2^{-3/2}\{ \frac{\eta_0}{3}-[\frac{\eta_0}{3}-\frac{1}{2}][\frac{n(0)}{n(t_0)}]^{1/2}\}^{-3/2}\\
&=2^{-\frac{3}{2}}\{ \frac{\eta_0}{3}[\frac{n(t_0)}{n(0)}]^{-\frac{2}{3}}-[\frac{\eta_0}{3}-\frac{1}{2}][\frac{n(t)}{n(0)}]^{-\frac{7}{6}}\}^{-\frac{3}{2}}.
\end{split}
\end{equation}
%%%%%%%%%%%%%%%%%%%%%%%%%%%%%%%%%%%%
$n(t)/n(0)\leq 1$. Ultimately, we can figure out how $F_\rho$ changes with $T(0)$. $F_\rho$ is not monotonic. The solution of its first derivation is
\begin{equation}\nonumber
    \left[ \frac{8\eta_0}{7(2\eta_0-3)}\right]^{-2}.
\end{equation}
The zero point is larger than 1 if $\eta_0>3.5$. In other word, $F_\rho$ becomes larger with lower temperature  as long as $\eta_0>3.5$.

%%%%%%%%%%%%%%%%%%%%%%%%%%%%%%%%%%%%

\subsection{Numerical simulation of the $\eta(t)$ for a maximum $\rho$}
We present the numerical simulation for the optimized $\rho$. The results show the optimized value of $\eta_0$ closes to $9/2$, which agrees with previous theoretical prediction too. In our three-body cooling, $\eta_0$ keeps the same without varying, which results in the simulated optimized $\eta_{max}(t)$ off from $9/2$. The big jump at 350 ms in the $T(0)=0.6 \mu$K experiment is because of the large cooling from $\eta_0=3$ to $\eta_0=7$. Ideally, if $L_3\propto e^{-\eta}$ is right all the time, the simulated optimized $\eta_{max}(t)$ will be the same tendency of the $T(0)=5.1\mu$K case and reach about $9/2$. But, the three-body cooling modifies the changing of $L_3$ and results in the numerical simulation off the original prediction.
%%%%%%%%%%%%%%%%%%%%%%%%%%%
\begin{figure}[htbp]
    \begin{center}
        \includegraphics[width=\columnwidth, angle=0]{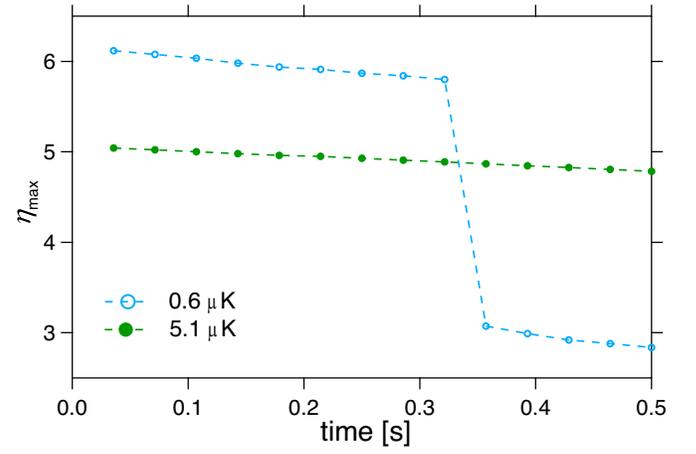}
        \caption{ Numerical simulation of the $\eta(t)$ for a maximum $\rho$. The green dots are for the $T(0)=5.1 \mu$ K case, and light blue open circles are for the $T(0)=0.6\mu$K.
        }    \label{p:numeriacl_CE}
    \end{center}
\end{figure}
%%%%%%%%%%%%%%%%%%%%%%%%%%%

\newpage
%-------------------------------------------------------------------------------------%

%-------------------------------------------------------------------------------------%

\end{document}